\documentclass[%
 reprint,
 amsmath,amssymb,
 aps,
]{revtex4-1}

\usepackage{graphicx}
\usepackage{dcolumn}
\usepackage{bm}
\usepackage{color}

\renewcommand{\vec}[1]{\ensuremath{\mathbf{#1}}}		

\newcommand{\kheat}{\ensuremath{k}}				
\newcommand{\cheat}{\ensuremath{c_{p}}}			
\newcommand{\lDebye}{\ensuremath{\lambda_\mathrm{D}}}		
\newcommand{\kB}{\ensuremath{k_\mathrm{B}}}			
\newcommand{\Eunit}{\ensuremath{\left( \frac{Q^4m\omega^2}{(4\pi\epsilon_0)^2}\right)^{1/3}}}		
\newcommand{\runit}{\ensuremath{\left( \frac{Q^2}{4\pi\epsilon_0 \cdot m\omega^2}\right)^{1/3}}}	

\begin{document}

\preprint{APS/123-QED}

\title{Heat Transport in Confined Strongly Coupled 2D Dust Clusters}

\author{Giedrius Kudelis}
 \affiliation{University of Birmingham, Edgbaston, Birmingham B15 2TT, United Kingdom}
\author{Hauke Thomsen}%
 \email{thomsen@theo-physik.uni-kiel.de}
\author{Michael Bonitz}
 \email{bonitz@theo-physik.uni-kiel.de}
\affiliation{%
 Institut f\"ur Theoretische Physik und Astrophysik\\
 Christian-Albrechts-Universit\"at zu Kiel, 24098 Kiel, Germany
}%

\date{\today}

\begin{abstract}
 Dusty plasmas are a model system for studying strong correlation. The dust grains' size of a few micro-meters and their characteristic oscillation frequency of a few hertz allows for an investigation of many particle effects on an atomic level. In this article, we model the heat transport through an axially confined 2D dust cluster from the center to the outside. The system behaves particularly interesting since heat is not only conducted within the dust component but also transfered to the neutral gas. Fitting the analytical solution to the obtained radial temperature profiles allows to determine the heat conductivity $\kheat$. The heat conductivity is found to be constant over a wide range of coupling strengths even including the phase transition from solid to liquid here, as it was also found in extended systems by V.~Nosenko et al. in 2008~\cite{PhysRevLett.100.025003}. 
\end{abstract}

\maketitle

\section{Introduction}
Strongly correlated systems are of high interest in modern physics. Strong correlations can emerge at very different physical regimes. In dusty plasmas, the high particle charges are responsible for the strong correlations, while ions in traps or laser plasmas can be strongly correlated due to the small particle distances~\cite{bonitz_complex_2010}.

Besides static properties, experiments and Molecular Dynamics simulations provide insight to dynamic properties. Particle transport (diffusion) is well investigated~\cite{PhysRevLett.103.195001,PhysRevLett.107.135003}. Also, the heat transport is is an important property that was experimentally investigated in extended systems~\cite{PhysRevLett.100.025003,PhysRevE.86.056403}. One question is how the strong correlations affect the transport properties.

In this article, we investigate the heat transport in a finite 2D cluster in an experiment-oriented simulation. After a brief presentation of the system model and the simulation technique in Sec.~\ref{sec:setup}, we present simulation results for different parameters in Sec.~\ref{sec:results}. The analysis focuses on the temperature of the dust particles' random motion and the radial temperature profiles under different heating conditions. These profiles are compared to the solution of the stationary heat transport equation for the 2D cluster in Sec.~\ref{sec:analytic}, which allows to derive the thermal conductivity $\kheat$ under certain assumptions. Finally, we discuss the results for $\kheat$ and the dependence of $\kheat$ on heating power, equilibrium coupling strength and screening parameter in Sec.~\ref{sec:discuss}.

\section{Experimental setup and simulation model\label{sec:setup}}
2D dust clusters are usually realized by using a lower electrode with a cavity. The plasma sheath which contains the dust grains then reflects the shape of the cavity. This shape allows for a horizontal confinement of the dust. In order to investigate heat transport, it is necessary to heat to the dust component in a restricted area. The laser manipulation technique~\cite{schablinski_laser_2012,thomsen_laser_2012} is well suited for this purpose. Several randomly moving laser spots accelerate dust particles in different directions due to the momentum transfer by the radiation pressure. A spatially inhomogeneous heating is achieved by restriction of the area that is scanned by laser beams.

The unheated $N$-particle system is described by the Hamiltonian 
\begin{equation}
 H = \sum_{i=1}^{N}\frac{\vec{p}_i^2}{2m}
    + \sum_{i=1}^{N} \frac{m \omega^2}{2} \vec{r}_i^2
    + \sum_{i<j} \frac{Q^2}{4\pi\epsilon_0 r_{ij}} e^{r_{ij}/\lDebye} \text{ ,}
\end{equation}
where $r_{ij}=|\vec{r}_j - \vec{r}_i|$ is the pair distance and $\lDebye$ is the Debye length. This model assumes that all particle are equal in mass $m$ and charge $Q$. Wake effects that occur in streaming plasmas are neglected, since a single layer of particles is subject to our studies. The horizontal interaction is adequately describe by a Yukawa potential~\cite{piel_dynamical_2002}. This Hamiltonian is transformed to a dimensionless form by introducing $t_0=1/\omega$, $l_0=\runit$ and $E_0=\Eunit$ as units for time, length and energy. The Debye screening is then characterized by the screening parameter $\kappa=l_0/\lDebye$.

While the dust component is treated exactly, electrons, ions and neutral are treated statistically in the Langevin model. In this model, the amplitude of the stochastic force is expressed in therms of the friction coefficient $\gamma$ between the dust particles and the neutral gas and the equilibrium temperature $T_\mathrm{eq}$. $T_\mathrm{eq}$ and $\gamma$ are input parameters to the simulation.  We use Langevin Molecular Dynamics simulations~\cite{PhysRevE.69.041107} to propagate the equations of motion for the dust component and obtain the trajectories. The heating lasers are included as space and time dependent forces~\cite{thomsen_laser_2012}
\begin{equation}
  \vec{f}_l(\vec{r},t) = \frac{P_0}{2\pi \sigma_x \sigma_y} \vec{e}_l \exp\left\lbrace
    - \frac{(x-x_l(t))^2}{2\sigma_x^2} - \frac{(y-y_l(t))^2}{2\sigma_y^2}
  \right\rbrace 	\text{ ,}
 \label{eq:laser_force}
\end{equation}
in this model. The index $l$ counts the laser. The force amplitude is given by $P_0$, $\sigma_{x/y}$ describe the anisotropic spot profile and $\vec{e}_l$ the a unit vector in beam direction. $\vec{r}_l(t)=(x_l(t),y_t(t))$ is the trajectory of the moving spot in the levitation plane. We use two pairs of laser spots, one pair in $x$ and one in $y$ direction. The lasers of each pair accelerate the particles in opposite direction, thus the average total momentum transfer vanishes. The laser trajectories are generated randomly. The spot $l$ moves across the levitation plane with a velocity $\vec{v}_l$. When it reaches a border of the heated window, the motion is reverted in this direction and a new speed is randomly chosen within an interval. This method was shown to allow for a controlled heating of the dust component without altering the properties of the plasma and the neutral gas~\cite{schablinski_laser_2012,thomsen_laser_2012}.

We use the Coulomb coupling parameter $\Gamma=Q^2/(4\pi\epsilon_0 l_0 \kB T)$~\cite{PhysRevE.72.026409,PhysRevA.35.3109} as ration of typical interaction energy and thermal energy to characterize the coupling strength. As characteristic particle distance, the unit length $l_0$ is used.

\section{Simulation results\label{sec:results}}
\begin{figure}[b]
\includegraphics{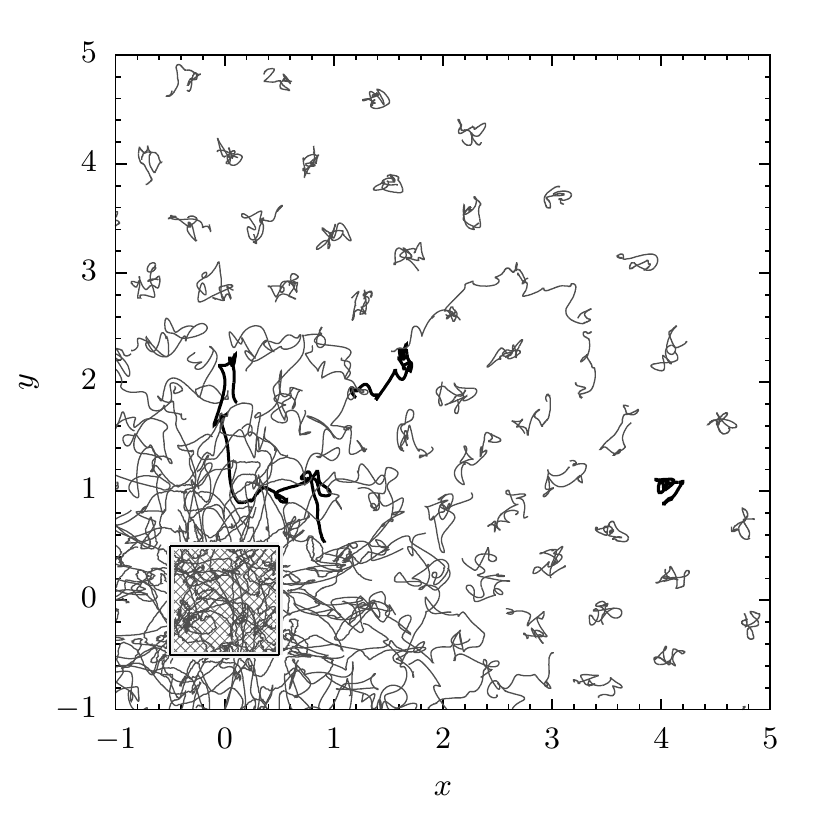}
\caption{\label{fig:traj} Trajectories in the first quadrant of a heated Yukawa cluster during $\Delta t=10 t_0$. Each one arbitrary particle is highlighted in the central, middle and outer region. The gray pattern around the origin indicates the array scanned by the laser spots.\\ Parameters: particle number $N=200$, screening parameter $\kappa=1$, equilibrium Coulomb coupling parameter $\Gamma_{eq}=200$, friction parameter $\gamma=0.5$ and heating power $P=90$.}
\end{figure}

Fig.~\ref{fig:traj} shows an example for the trajectories in a cluster with $N=200$ particles, with $\kappa=1$ and $\Gamma_\mathrm{eq}=200$. The four laser spots move within a square with a side length $a=1$ in units of $l_0$.  While the motion is restricted to a small region around the equilibrium position for particles at the cluster's edge, particles in the central region are less localized. 

\begin{figure}
  \includegraphics{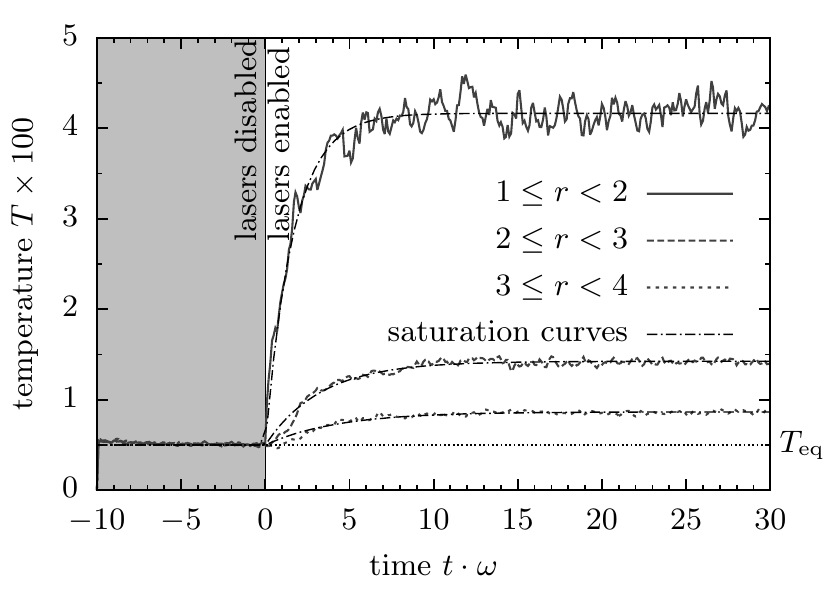}
  \caption{\label{fig:Tvst}Time evolution of the local temperature in the simulation for different radii. The lasers were enabled at $t=0$. While the temperature at the inside saturates on a time scale $\tau=1.7$, the relaxation is slower for the outer particles with $\tau=4.3$. The temperatures where average over 40 independent simulations.\\
  Parameters: $N=200$, $\kappa=1$, $\Gamma_\mathrm{eq}=200$, $\gamma=0.5$, $P=50$}
\end{figure}

\begin{figure}
\includegraphics{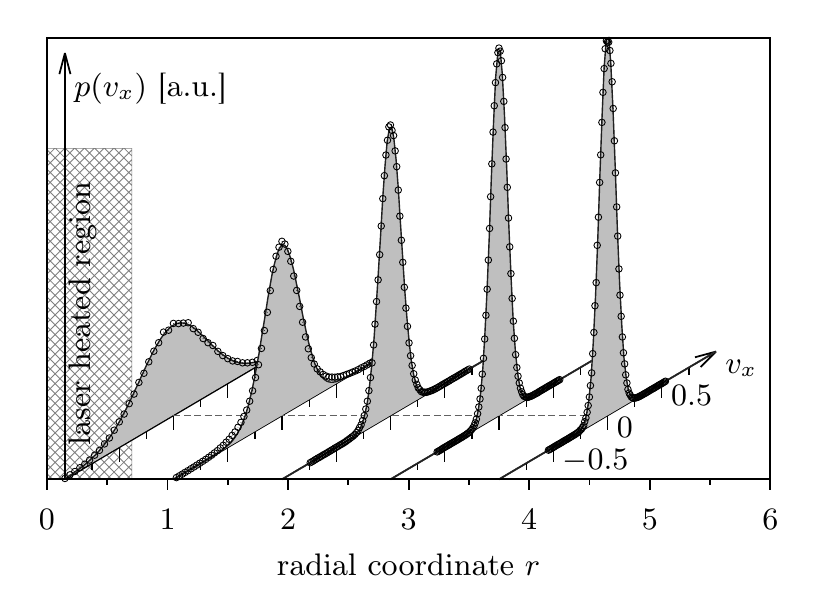}
\caption{\label{fig:velocity_histo} Velocity distribution $p(v_x)$ of particles within concentric rings ($N=200$, moderate heating power $P=40$, equilibrium coupling strength $\Gamma_{eq}=200$). The data points are well fit by Gaussians (filled curves) with decreasing width $\sigma_v=\sqrt{\kB T(r)/m}$. The velocity distribution in $y$-direction $p(v_y)$ coincides with $p(v_x)$ and is omitted for this overview. The heated area is indicated by the gray pattern.}
\end{figure}

The trajectories in Fig.~\ref{fig:traj} show that the inner particles move greater distances within the same time than the outer ones. In order to quantify the associated temperature gradient, we calculate the radial temperature profile. Therefore, the simulation space is divided into circular rings of width $\Delta r$. The width has to be chosen small enough to obtain a sufficient radial resolution and large enough to obtain a sufficient number of data points per ring. $\Delta r= 0.3$ in units of the trap length turned out as a proper choice. The temperature for the ring is calculated as $\kB T= m \left( \langle \vec{v}^2 \rangle - \langle \vec{v} \rangle^2 \right)$ by taking the average over the whole simulation after an equilibriation phase of typically $t_\mathrm{eq}=20\omega^{-1}$. The temperature profile reaches a steady state a few plasma periods $\omega^{-1}$ after powering the heating lasers in the simulation. The time resolved temperatures for three different radii are shown in Fig.~\ref{fig:Tvst}. Since the dust cluster is trapped, the collective flow velocity $\langle \vec{v} \rangle=0$ vanishes. Figure~\ref{fig:velocity_histo} shows the velocity distributions of different rings from the center to the outside of the cluster. The distributions are Maxwellian for the inner rings as well as for the outer rings. The width of $p(v)$ is large in the center and decreases monotonically towards the cluster's border. That means, the temperature decreases with the radius, as expected.

\subsection{Analytical model\label{sec:analytic}}
In this section, we present an analytical model that can describe the radial dependence of the temperature. Our goal is to describe the dust temperature outside the central heated region.
Therefore, we consider a incompressible fluid model and add the thermal coupling to the background gas with temperature $T_\mathrm{eq}$ in the heat transport equation~\cite{Landau_Lifschitz_VI}, 
\begin{equation}
  \cheat n \left(\frac{\partial T}{\partial t} + \vec{v} \vec{\nabla} T \right) 
	= \mathrm{div}\left( \kheat \vec{\nabla} T  \right)
 	  + S_\mathrm{v}
	  -2\gamma n (T-T_\mathrm{eq}) \text{ .}
  \label{eq:heat_trans_time}
\end{equation}
On the right hand side of Eq.~(\ref{eq:heat_trans_time}), we find the convective time derivative of the temperature. The right hand side consists of the thermal conduction, the shear heating as function of the spatial derivative of the velocity vector
\begin{equation}
 S_\mathrm{v} = \frac{\eta}{2}\left(\frac{\partial v_i}{\partial x_k} + \frac{\partial v_k}{\partial x_i} \right)^{2} \text{ ,}
 \label{eq:shear_heating}
\end{equation}
where the indices $i$ and $j$ take values $1,2$ in two dimension and the squaring implies summation over $i$ and $k$. 
The last term describes the losses to the neutral gas. If we wanted to describe the central region where the laser heating takes place, we would have to introduce another source term. When we consider the dust cluster long after equilibration, the temperature is constant and the partial time derivative $\frac{\partial T}{\partial t}$ vanishes. 

\begin{figure}
 \includegraphics{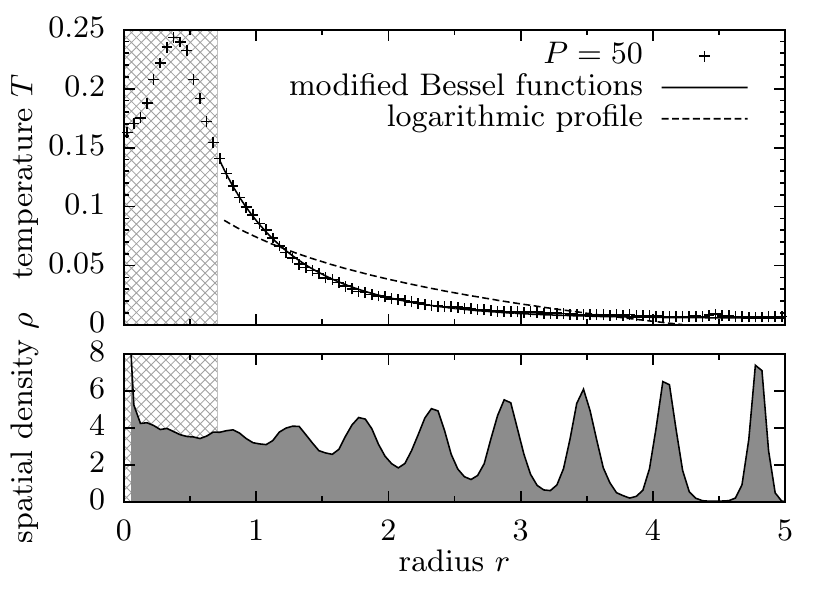}
  \caption{\label{fig:k_fit_single} The simulation data (symbols in upper plot, $N=200$, $\kappa=1$, $T_\mathrm{eq}=0.005$, $\gamma=0.5$) are fit by both a logarithmic temperature profile $T(r)= c_1 \ln(r/r_0)$ and by modified Bessel functions, Eq.~(\ref{eq:solution_bessel}). The latter fits the data well while the former solution does not. The central region was excluded from the fits since the power input by the lasers takes place in this region. The spatial density (lower plot) is alsmost constant in the central region where most of the heat loss to the neutral gas takes place.}
\end{figure}

The \textit{stationary} heat transport equation\footnote{We added the Boltzmann constant $\kB$ in Eq.~(\ref{eq:heat_trans_time}) to allow temperatures in Kelvin. In the heat transport term, $\kB$ is absorbed by the thermal conductivity $\kheat$. Furthermore, we use the areal particle density $n$ instead of the mass density $\rho$.} reads
\begin{equation}
 \cheat n\vec{v} \cdot \vec{\nabla} T \kB = \mathrm{div}(\kheat \vec{\nabla}T) - 2\gamma n(T-T_\mathrm{eq})\kB + S_\mathrm{v} \text{ ,}
 \label{eq:heat_trans_full}
\end{equation}
where $\cheat$ is the specific heat, $n$ the spatial number density and $\vec{v}$ is the flow velocity~\cite{PhysRevLett.100.025003}. While the equilibrium temperature $T_\mathrm{eq}$ and the friction coefficient $\gamma$ are input parameters to the simulation, the heat conductivity $\kheat$ is unknown. We further neglect the convective heat transport on the left side and heat input by viscous conversion of energy form a shear flow $S_\mathrm{v}$. This is motivated by the observation that no shear flow is observed in the simulation data. We check the validity of the assumptions later by comparing the temperature profiles in the simulations with the analytic solution. The reduced heat transport equation then reads
\begin{equation}
 \mathrm{div}(\kheat \vec{\nabla}T) = 2\gamma n \kB (T-T_\mathrm{eq}) \text{ .}
  \label{eq:heat_trans_reduced}
\end{equation}
Due to the symmetry, we expect the temperature to depend on the radial coordinate $r$ only, and rewrite Eq.~(\ref{eq:heat_trans_reduced}) in polar coordinates as 
\begin{equation}
 \frac{1}{r}\frac{d}{d r}\left(r \kheat \frac{d T}{d r} \right)
  = 2 \gamma n \kB (T - T_\mathrm{eq}) \text{ .}
  \label{eq:heat_trans_reduced_radial}
\end{equation}
When losses to the neutral gas are neglected ($\gamma=0$) and the heat conductivity $\kheat$ is independent of $r$, the solution of Eq.~(\ref{eq:heat_trans_reduced_radial}) is a logarithmic temperature profile $T(r)=c_1 \ln{(r/r_0)}$, with the integration constants $r_0$ and $c_1$. As shown in Fig.~\ref{fig:k_fit_single}, this solution does not reproduce the velocity profiles from the simulation. The temperature in the simulations decays much faster than the logarithmic profile.
The logarithmic solution would decay indefinitely, while we expect the system to be always warmer than the surrounding gas due to heating applied.
Hence, losses to the neutral gas by friction are important for the spatial heat distribution. These losses are responsible for the faster decay. The general solution of Eq.~(\ref{eq:heat_trans_reduced_radial}) with a heat conductivity $\kheat$ independent of $r$ is readily found and given by modified Bessel functions of first kind $I_0$ and second kind $K_0$,
\begin{equation}
 T - T_\mathrm{eq} = A K_0(\sqrt{b} r) + B I_0(\sqrt{b} r) \text{ ,}
 \label{eq:solution_bessel}
\end{equation}
where the abbreviation $b=2\gamma n \kB / \kheat$ is used. Since the dust cluster has a finite radial extension $R$, no heat can be transferred further outside. A heat flow equal zero at $r=R$ is ensured in the solution, when the temperature gradient vanishes. When Eq.~(\ref{eq:solution_bessel}) is differentiated, we find the second coefficient $B=A K_1(\sqrt{b}R) / I_1(\sqrt{b}R)$.

The solution Eq.~(\ref{eq:solution_bessel}) then has two remaining free parameters $A$ and $b$ which depend on the details of the laser heating, in particular, the heating power (see below). 
Since the heating is not included in Eq.~(\ref{eq:heat_trans_reduced_radial}) we determine these parameters by fitting the function to the simulation data.
The central region where this input takes place is excluded from the fit. While the amplitude $A$ is connected  to the total power input, the thermal conductivity $\kheat$ is contained in $b$. 
The total heat that is transfered to the neutral gas outside the heated region ($R_i$) is calculated by integration of $dP_\mathrm{heat} =2 n \gamma \kB (T-T_\mathrm{eq}) dA$,
\begin{align}
 P_\mathrm{heat} &= 2\pi \int_{R_i}^{R}{rdr\ }2 n \gamma \kB (T(r)-T_\mathrm{eq})
  \label{eq:Power_01} \\
 &= 4\pi \bar{n} \gamma \kB \frac{A\cdot R_i}{\sqrt{b}}\left(K_1(\sqrt{b}R_i)-I_1(\sqrt{b}R_i)\frac{K_1(\sqrt{b}R)}{I_1(\sqrt{b}R)} \right) \text{ .} \nonumber
\end{align}
Since the $B/A=K_1(\sqrt{b}R)/I_1(\sqrt{b}R)$ decays fast, the second term can be neglected for clusters that are not to small. The ratio is $B/A < 2\cdot10^{-5}$ for $\sqrt{b}R\geq 5$ while it is of the order of $1$ at the inner radius with $\sqrt{b}R\approx 1$. The total heating power is hence approxmimatly given by
\begin{align}
  P_\mathrm{heat} &\approx 4\pi \bar{n} \gamma \frac{A\cdot R_i}{\sqrt{b}} K_1\left(\sqrt{b} R_i\right) \text{ .}
  \label{eq:Power_approx}
\end{align}
The dependence on the inner radius $R_i$ remains, since Eq.~(\ref{eq:Power_approx}) is the heat loss to the neutral outside the central region only and hence, depends on the size of this excluded region.

Eq.~(\ref{eq:Power_01}) establishes the relation of the two free parameters to the input power P.
Since the radial temperature profile is well fit by the modified Bessel functions as solution of Eq.~(\ref{eq:heat_trans_reduced_radial}), see Figs.~\ref{fig:k_fit_single}, \ref{fig:k_fit_Gamma} and \ref{fig:k_fit_Power}, we conclude that Eq.~(\ref{eq:heat_trans_reduced_radial}) correctly describes the physics.

\subsection{Influence of equilibrium temperature $T_\mathrm{eq}$ and heating power $P$}
In this section, we investigate the influence of $T_\mathrm{eq}$ and $P$ on the thermal conductivity. A central question is, if the heat conduction changes during the transition between the solid-like and liquid-like regimes. The melting point of a macroscopic 2D Coulomb ($\kappa=0$) system is well known as $\Gamma=137$~\cite{PhysRevE.72.026409}. Due to finite size effects and finite screening parameter $\kappa$, do not expect a sharp transition temperature but a transition range~\cite{PhysRevE.56.4671} and choose the invested range of $\Gamma$ accordingly.  

\begin{figure}
 \includegraphics{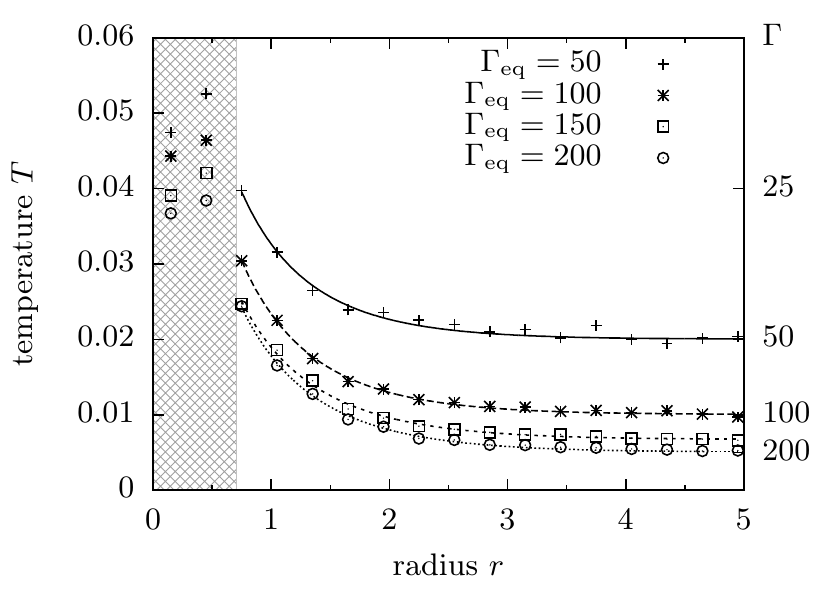}
  \caption{\label{fig:k_fit_Gamma}Simulation data (symbols) and fits by modified Bessel functions (lines) for different equilibrium temperatures and constant heating power $P=20$. The central region was always excluded. On the right axis, the temperature is translated into a local coupling strength.}
\end{figure}

\begin{figure}
 \includegraphics{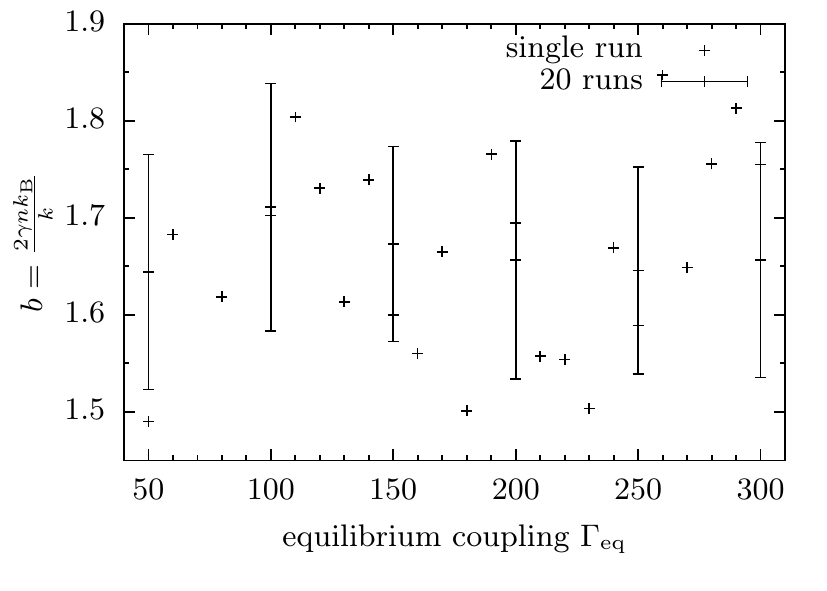}
 \caption{\label{fig:results_b_single}Values of the parameter $b$ for different equilibrium temperatures $\Gamma_\mathrm{eq}$. Each symbol represents a single simulation. The error bars show the variance of the single estimates of $b$.\\
 Parameters: $N=200$, $\kappa=1$, $\gamma=0.5$, $P=100$}
\end{figure}

\begin{figure}
 \includegraphics{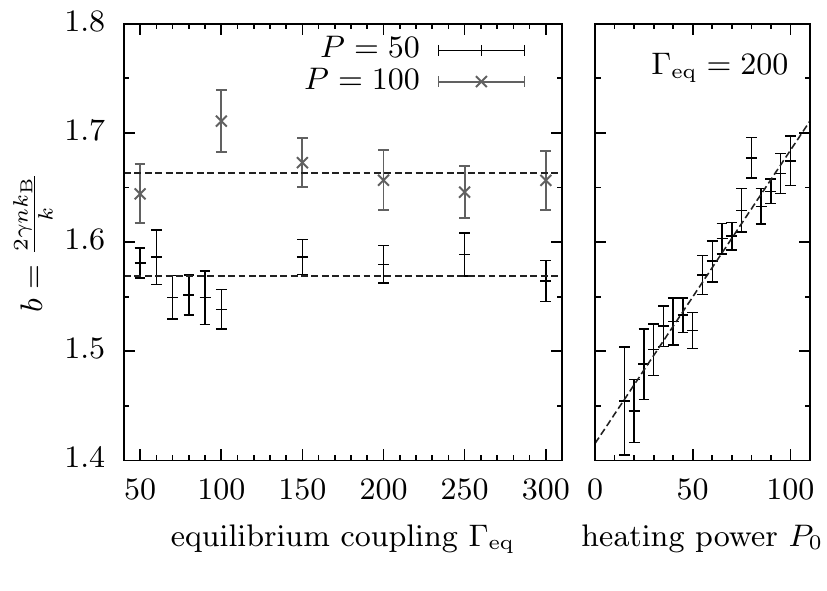}
 \caption{\label{fig:results_b}Values of the parameter $b$ for different equilibrium temperatures of the Langevin thermostat (left) and for different heating powers (right). Each symbol represents the average from 20 independent simulations with the variance of this average as error bar. The horizontal dashed lines indicate the mean value of $\bar{b}=1.57 \pm 0.01$ for $P=50$ and $\bar{b}=1.66 \pm 0.01$ for $P=100$ in dimensionless units.
 The power dependence is fit by a linear dependence $b=b_0+m \cdot P$ with $b_0=1.42 \pm 0.01$ and  $m=2.68 \pm 0.18 \times 10^{-3}$. \\
 Parameters: $N=200$, $\kappa=1$, $\gamma=0.5$}
\end{figure}

As a first parameter scan, the equilibrium temperature of the Langevin thermostat is varied at constant heating power. The radial temperature profiles for four different values of $T_\mathrm{eq}$ are shown in Fig.~\ref{fig:k_fit_Gamma}. We used a comparatively small laser power for this plot, since the background temperatures are of the same order of magnitude as the central temperatures in that case. The temperature approaches $T_\mathrm{eq}$ towards the cluster boundary in all four simulations. Varying $T_\mathrm{eq}$ has also an effect on the temperature of the heated particles in the center. The temperature is increased by an increase of $T_\mathrm{eq}$. This observation is not surprising, since heat input by collisions with the neutral gas (i.e. via the stochastic force) and heat input by the laser force add up. While the amplitude of the temperature increases with $T_\mathrm{eq}$ the parameter $b$ remains constant. However, the dispersion of the single measurements is large as seen in Fig.~\ref{fig:results_b_single}. In order to estimate the error of $b$, we performed a series of 20 simulation for each value $\Gamma_\mathrm{eq}=50,\hdots,300$. Each run uses the same parameters but different random numbers and initial particle positions. Mean value and standard deviation of $b$ are also plotted in Fig.~\ref{fig:results_b}. Within the accuracy, we find that $b=1.57\pm0.01$ ($P=50$) and $b=1.66\pm0.01$ ($P=100$) is constant when $T_\mathrm{eq}$ is varied. A series of 20 simulations was also performed with moderate coupling $\Gamma_\mathrm{eq}=1$ and $P=100$. The density profile becomes wide at the cluster boundary but is only changed weakly at the inside where most of the heat is transfered and lost to the neutral gas. With that, the result for $b=1.5\pm0.7$ indicates the independence of thermal conductivity and equilibrium temperature over a wide range. 

\begin{figure}
 \includegraphics{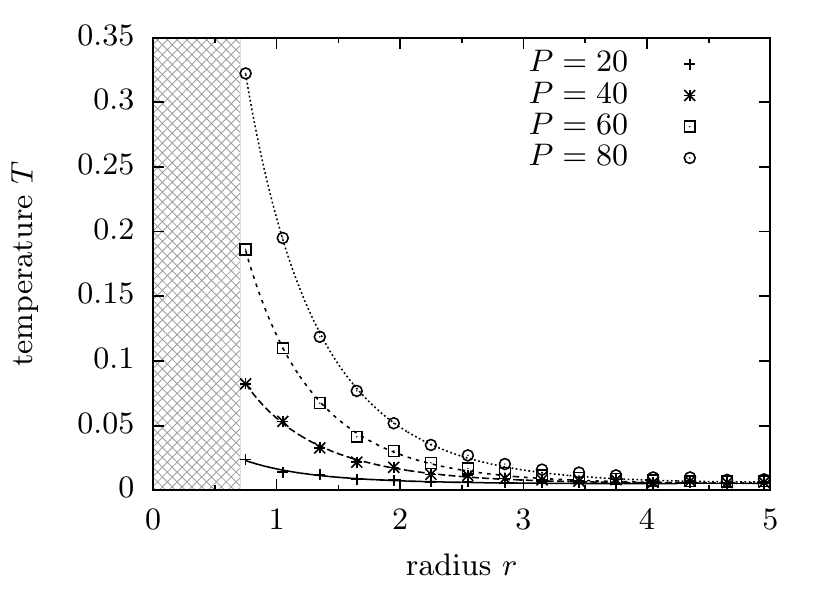}
  \caption{\label{fig:k_fit_Power}Simulation data (points) and fits by modified Bessel functions (lines) for different heating powers. The central region was always excluded. The equilibrium temperatures corresponds to $\Gamma_\mathrm{eq}=200$ in all cases.}
\end{figure}

In a second parameter scan, the laser power is varied while the equilibrium temperature is fixed corresponding to $\Gamma_\mathrm{eq}=200$. The temperature profiles for four exemplary powers are shown in Fig.~\ref{fig:k_fit_Power}. The higher power input results in an increased temperature in the cluster inside. For small powers, the equilibrium temperature is reached at the cluster's outer radius. For higher powers, the temperature of the outer particles is about twice as high as the equilibrium temperature. Interestingly, the parameter $b$ shows a linear dependence on the heating power within the accuracy of the estimation, see right side of Fig.~\ref{fig:results_b}.  

\section{Discussion of the results\label{sec:discuss}}
In this section, we give an interpretation of the simulation results and compare them with experimental results for bulk systems.
The parameter $b$ as introduced in Eq.~(\ref{eq:solution_bessel}) has the dimension of an inverse length squared. Hence, $L=1/\sqrt{b}$ can be interpreted as an \textit{characteristic length} for the heat transport. For arguments greater than $2$, the modified bessel function $K_1$ is well approximated by an exponential decay. The temperature difference $T(r)-T_\mathrm{eq}$ drops by a factor $\approx 2.5$ each $L$ in radial direction.  The mean value of $\bar{b}=1.66$ from the simulations for different equilibrium temperatures, Fig.~\ref{fig:results_b}, translates into $L=0.78$ which is approximate the average inter particle distance.
Another quantity which is often calculated in experiments~\cite{PhysRevLett.100.025003, PhysRevLett.95.025003, PhysRevE.75.026403} is the thermal diffusivity $\chi=k/(n \cheat)$, where $\cheat$ is the specific heat capacity. The $N$ particle system has $2N$ spatial degrees of freedom, where the common rotation is not connected with potential energy and $2N$ momentum degrees of freedom. The specific heat is approximated as $c_P=(4N-1)/2N\approx 2$. Using the value of $c_P=2$, we obtain a thermal diffusivity of 
\begin{equation}
 \chi=\frac{k}{n \cheat}= \frac{2\gamma n \kB}{b n \cheat} = \frac{\gamma \kB}{b} \approx 4.4 mm^2/s \text{ .}
  \label{eq:thermal_diffusivity}
\end{equation}
The conversion from dimensionless units into SI units was done for melamin particles with $d_p=6\mu m$ diameter, a charge of $Q=10,000 e$, a trap frequency $\omega=5.5s^{-1}$ and a friction frequency $\gamma=\omega/2$ as typical plasma parameters. This result is in reasonable agreement with experimental findings for $\chi\approx9 mm^2/s$ and $\chi_\mathrm{tr}\approx30 mm^2/s$ (transversal), $\chi_\mathrm{l}\approx50 mm^2/s$ (longitudinal) for 2D complex plasmas \cite{PhysRevLett.100.025003, PhysRevLett.95.025003} and $\chi\approx1 mm/s$ in a liquid 3D complex plasma~\cite{PhysRevE.75.026403}.

We have shown that the radial heat transport in a confined 2D Yukawa cluster is well described by a fluid model. In this model, we could find an analytical form of the radial temperature profile. Using this model, Langevin Molecular Dynamics results show that the heat conductivity is constant over the investigated temperature range. This range includes the transition region between liquid-like and solid-like cluster. 

\textbf{Acknowledgements:} This work is supported by the Deutsche Forschungsgemeinschaft via SFB-TR24 (project A5), by the German Academic Exchange Service via the RISE program and a grant for computing time at the HLRN.

\nocite{*}

\providecommand{\noopsort}[1]{}\providecommand{\singleletter}[1]{#1}%

\end{document}